\documentclass[aps,twocolumn,pra,superscriptaddress]{revtex4-2}
\usepackage{amsfonts}
\usepackage[final,hiresbb]{graphicx}
\usepackage{amsmath}
\usepackage{amssymb}
\usepackage{amstext}
\usepackage{color}
\usepackage{braket}

\usepackage{subfigure}
\usepackage{appendix}

\definecolor{Green}{RGB}{0,204,102}
\definecolor{Purple}{RGB}{102,0,255}
\definecolor{Blue}{RGB}{51,153,255}
\definecolor{Red}{RGB}{255,010,010}

\newcommand{\delg}{\Delta_{\scriptscriptstyle \rm Gouy}}
\newcommand{\dnat}{d_{ \scriptstyle \rm{nat}}}
\newcommand{\fnat}{f_{ \scriptstyle \rm{nat}}}
\newcommand{\HG}{H\!G}
\newcommand{\phips}{\phi}
\newcommand{\thetaps}{\theta}

\newcommand{\phii}{\phi_{ \scriptstyle \rm{i}}}

\newcommand{\xps}{x_{\scriptscriptstyle \!S\!O\!M}}
\newcommand{\yps}{y_{\scriptscriptstyle \!S\!O\!M}}
\newcommand{\zps}{z_{\scriptscriptstyle \!S\!O\!M}}
\newcommand{\zl}{z_\ell}
\newcommand{\zla}{z_{\ell 1}}
\newcommand{\zlb}{z_{\ell 2}}

\newcommand{\zRp}{z_{\scriptscriptstyle Rp}}
\newcommand{\zRa}{z_{\scriptscriptstyle Ra}}
\newcommand{\zrel}{z_{\scriptscriptstyle \rm rel}}
\newcommand{\zrela}{z_{\scriptscriptstyle \rm rel,1}}
\newcommand{\zrelb}{z_{\scriptscriptstyle \rm rel,2}}

\begin{document}
	\title{The Anatomy of Geometric Phase for an Optical Vortex Transiting a Lens}
	
	\author{Mark T. Lusk}
	\email{mlusk@mines.edu}
	\affiliation{Department of Physics, Colorado School of Mines, Golden, CO 80401, USA}
	
	\author{Andrew A. Voitiv}
	\affiliation{Department of Physics and Astronomy, University of Denver, 2112 E. Wesley Avenue, Denver, CO 80208, USA}
	
	\author{Chuanzhou Zhu}
	\affiliation{Department of Physics, Colorado School of Mines, Golden, CO 80401, USA}
	
	\author{Mark E. Siemens}
	\email{Mark.Siemens@du.edu}
	\affiliation{Department of Physics and Astronomy, University of Denver, 2112 E. Wesley Avenue, Denver, CO 80208, USA}

	\begin{abstract}
		We present an analytical means of quantifying the fractional accumulation of geometric phase for an optical vortex transiting a cylindrical lens. The standard fiber bundle of a Sphere of Modes is endowed with a Supplementary Product Space at each point so that the beam waists and their positions can be explicitly tracked as functions of lens transit fraction. The method is applied to quantify the accumulation of geometric phase across a single lens as a function of initial state and lens position within the beam. It can be readily applied to a series of lenses as well. 
	\end{abstract}
	\maketitle

	\section{Introduction}

The interplay of geometry and information is at the heart of our quest to understand Nature, a perspective that has materialized from the merging of two fronts: our evolving understanding that general relativity~\cite{Carroll2004} and quantum field theory~\cite{Quigg2013} are based on the geometric nature of gauge fields; and, that the interpretation of any measurement is an application of information theory~\cite{Nielsen2010}. Essentially all of the scientific and technological progress of the last century might be summed up as milestones in our understanding of the relationship between these two concepts. 

A particularly rich manifestation of this interplay is the geometric holonomys exhibited by fields as they evolve~\cite{Wilczek1989, Berry1990}. In simple settings, there is a geometric phase that is just the angle between initial and final orientations of a vector field when parallel transported in a closed loop---e.g. the holonomy of Foucault's pendulum~\cite{Oprea1995}. At the other extreme, the holonomy is characterized by non-Abelian gauge fields that form the basis of the standard model~\cite{Quigg2013} and also underlie current research in topological quantum computing~\cite{Nayak2008}.

Classical and quantum optics comprise an accessible common ground for the study and harnessing of geometric information. Few-photon states embrace higher-dimensional parameter spaces and admit the study of entanglement in geometric phase, while classical optics can still exhibit non-Abelian behavior through synthetic gauge fields~\cite{Iadecola2016,Shen2020,Chen2021}. With an eye towards these richer settings, the transit of classical electromagnetic beams through linear optical elements offers a particularly illuminating arena for elucidating how and why geometric phase accumulates. There 3D electromagnetic modes can be viewed as 2D fields with the propagation axis interpreted as time.


Laser beams composed of linear combinations of Laguerre-Gaussian (LG) modes may accumulate geometric phase as they transit linear optical elements~\cite{vanenk1993,Padgett1999}. The associated parameter space can be projected onto a Sphere of Modes (SOM) (sometimes referred to as the modal or optical Poincar\'e Sphere) so that the geometry of SU(2) underlies such geometric information. Experimental progress in this area has focused on the use of optical elements that collectively trace out closed circuits composed of geodesic arcs~\cite{Galvez2003}.  

An explicit determination of non-geodesic trajectories has not been undertaken in the SOM setting, and the calculation of geometric phase for non-cyclic processes~\cite{Samuel1988} has yet to be considered as well. Previous calculations of geometric phase lean heavily on the results from the more studied setting in which it is polarization that evolves~\cite{Courtial1999a}. The analogy is limited, though, since projection onto the SOM necessarily disregards information about basis vectors which are themselves evolving~\cite{vanenk1993}. As result, the way in which geometric phase is actually accumulated is obscured by drawing on the polarization analogy. 

\begin{figure}[t]
	\begin{center}
		\includegraphics[width=\linewidth]{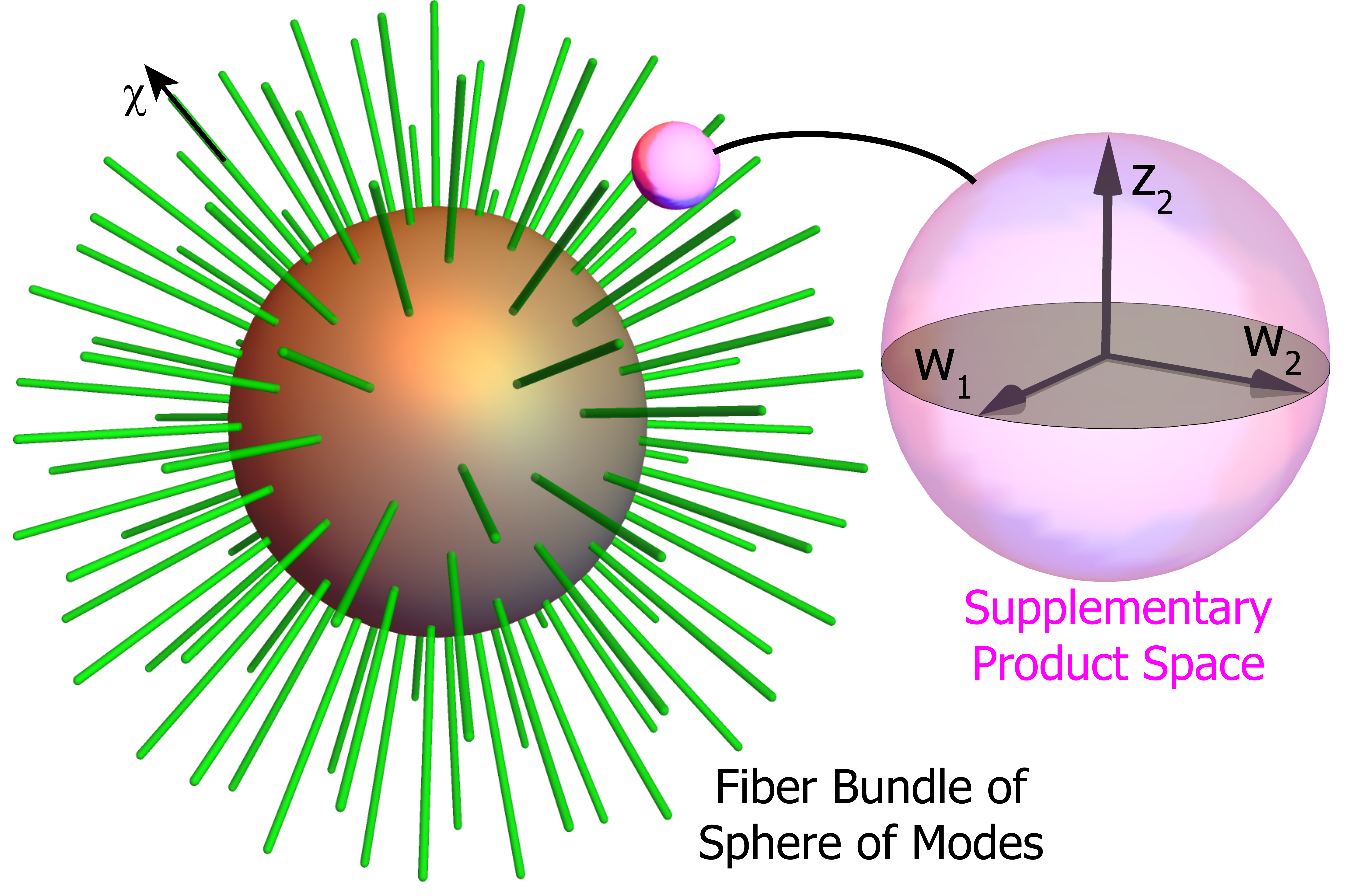}
	\end{center}
	\caption{ \emph{Fiber Space}. The Sphere of Modes (brown) is the projection of a 3D product space (pink),  $\{w_1, w_2, z_2 \}$, onto its associated fiber (green) followed by a projection of the fiber position, $\chi$, down to its base. Subscripts "1" and "2" refer to principle axes of an elliptical beam cross-section. Here $z=0$ is defined as the position of waist $w_1$.} 
	\label{Fiber_Space_5}
\end{figure}

These issues are addressed in the present work, and one immediate dividend of the approach is that it allows geometric phase to be evaluated as a function of lens transit fraction---explicitly showing that geometric phase accumulation arises from light-matter interactions within the lens rather than free-space propagation after the lens. The SOM is extended to a fiber bundle, shown in Fig. \ref{Fiber_Space_5}, so that the total phase of the beam can be tracked. Optical elements are decomposed into differential constituents allowing phase to be calculated as a function of transit fraction. Each point on the fibers is additionally endowed with a Supplementary Product Space (SPS) so that the beam waists and their relative positions can also be explicitly traced as beams move through optical elements.

%
%
%

\begin{figure}[t]
	\begin{center}
		\includegraphics[width=\linewidth]{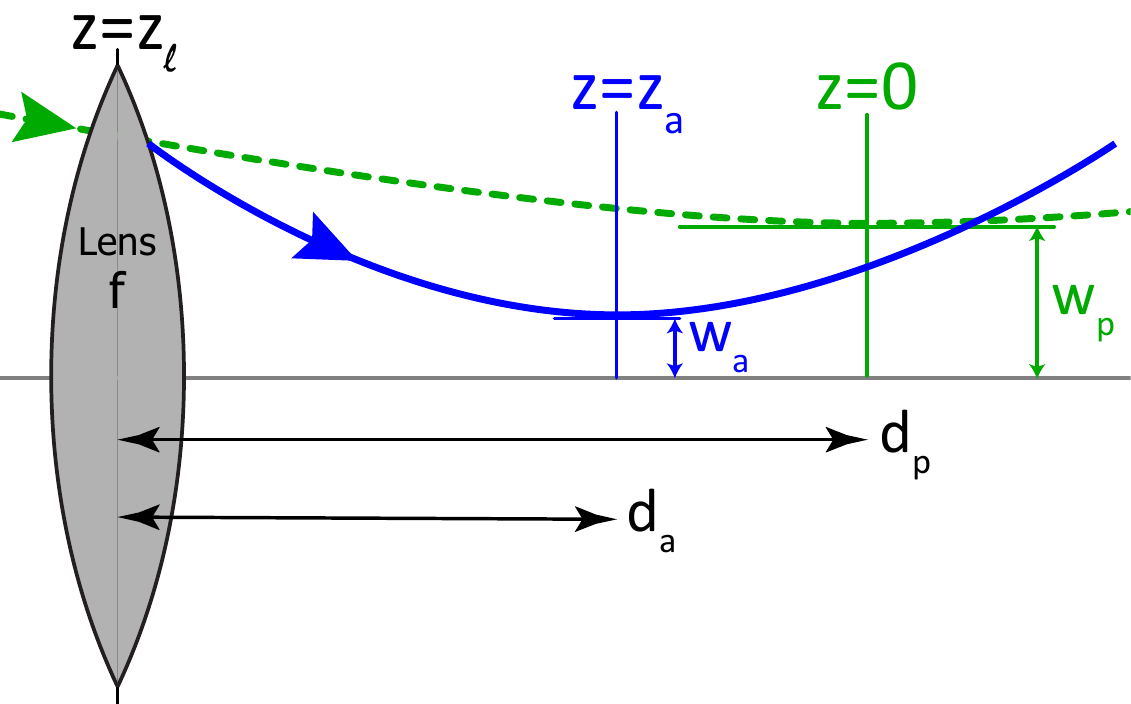}
	\end{center}
	\caption{ \emph{Lens Transit}. A thin, converging, cylindrical lens changes the waist and waist position along its active axis. Subscripts "\emph{p}" and "\emph{a}" refer to the passive and active axes of the cylindrical lens.} 
	\label{Lens_Focusing_Schematic_2}
\end{figure}

The method is applied to the transit of a beam composed of $LG_{0,+1}$ and $LG_{0,-1}$ modes---i.e. with zero radial mode number and opposite charges---with an optical vortex through a single cylindrical lens placed at a specified position relative to the beam waist, as shown in Fig. \ref{Lens_Focusing_Schematic_2}. The full trajectory across the SOM and through the SPS is calculated as a function of transit fraction. The geometric phase is then calculated for each lens using two distinct approaches using a geometric notion of distant parallelism that does not require that the SOM trajectory be closed.

%
%
%
%

\section{State Space for Tilted Vortices}

Under a paraxial approximation for the monochromatic electromagnetic vector potential~\cite{Lax1975}, free-space electrodynamic wave propagation is governed by a 2D Schr\"odinger equation~\cite{Lax1975} with axial position playing the role of time:
\begin{equation}\label{paraxial}
	i \partial_z u = -\frac{1}{2k} \nabla^2_\perp u.
\end{equation}

This governs the motion of a tilted Gaussian vortex~\cite{Andersen2021} whose transit through cylindrical lenses is most easily carried out by expressing it in terms of linear combinations of products of  normalized 1D Hermite-Gaussian (HG) modes:
\begin{eqnarray}\label{HG1D}
	u_{m}(x,z) &=& A_{p,m}(z) e^{-\frac{x^2}{B_p^2(z)}} \nonumber \\ 
	&& \mathbb{H}_m\biggl(\frac{x \sqrt{2}}{B_p(z)}\biggr) e^{i\frac{k x^2}{2 R_p(z)}}e^{-i \psi_p(z,m)}.
\end{eqnarray}
Here
\begin{equation}\label{norm}
	A_{p,m}(z) := \frac{1}{(2\pi)^{1/4} \sqrt{2^{m-1}m! B_p(z)} }
\end{equation}
is a normalization factor, $\mathbb{H}_m$ is a Hermite polynomial of order m, $B_p(z)$ is the beam radius, $R_p(z)$ is the radius of curvature of the beam, and $\psi_p(z,m)$ is the Gouy phase:
\begin{eqnarray}\label{radiusgouy}
	B_p(z) &=& w_p\sqrt{1+\frac{(z-z_p)^2}{\zRp^2}} \nonumber \\
	R_p(z) &=& (z-z_p)\biggl(1 + \frac{\zRp^2}{(z-z_p)^2}\biggr) \\
	\psi_p(z,m) &=& \biggl(m+\frac{1}{2}\biggr){\rm tan}^{-1}\biggl(\frac{z-z_p}{\zRp}\biggr) . \nonumber 
\end{eqnarray}
The beam waist, $w_p$, is located at $z_p=0$, and $\zRp := k w_p^2/2$ is the Rayleigh length. Subscripted "\emph{p}" labels indicate the assignment of the x-axis as the \emph{passive} axis of the cylindrical lenses that will be considered shortly. For prescribed wavenumber, $k$, the mode of Eq. \ref{HG1D} can be expressed as a ket,
\begin{equation}\label{ket}
	u_{m}(x,z) \equiv \braket{x,z|m^{w_p,z_p}},
\end{equation}
where the superscripts completely characterize the 1D beam by identifying its waist and corresponding axial position. Although bra-ket notation is more frequently associated with quantum mechanics, the features that make it convenient there hold for any Hilbert space with complex scalars, including the present setting.

Two-dimensional modes can then be constructed in which HG modes are defined on both x and y axes:
\begin{equation}\label{ket2}
	u_{m}(x,z)u_{n}(y,z) \equiv	\braket{x,y,z|m^{w_p,z_p}n^{w_a,z_a}}.
\end{equation}
Note that the y-dependent HG mode is generalized to allow for an \emph{active} "\emph{a}" waist and waist position that differs from the mode that depends on x. With these 2D modes as basis vectors, general first-order modes can be expressed in terms of the components of a spinor, $\{v_1, v_2 \} \in SU(2)$:
\begin{equation}\label{Jones1}
	\ket{u_{\rm vortex}} = v_1 \ket{1^{w_p,z_p}0^{w_a,z_a}} + v_2 \ket{0^{w_p,z_p}1^{w_a,z_a}} 
\end{equation}
with
\begin{eqnarray}\label{Jones2}
	v_1 &=& e^{i \chi} \cos(\theta/2)e^{-i \phi/2} \nonumber \\
	v_2  &=& e^{i \chi} \sin(\theta/2)e^{i \phi/2} .
\end{eqnarray}
%
%
Here $\phi\in[0,2\pi]$ and $\theta\in[0,\pi]$. It is important to keep in mind, though, that the spinor itself is an incomplete description of the state because all  information on the beam waists and their posisitions has been flensed.

\subsection{Sphere of Modes Fiber Bundle and Its Supplementary Product Space}

The spinor can be projected onto the SOM using a Hopf map to obtain the following SOM coordinates for the requisite Cartesian coordinates:

\begin{eqnarray}\label{PScart}
	\xps &=& |v_1|^2 - |v_2|^2 = \cos\theta  \nonumber \\
	\yps &=& 2 {\rm Re} (v_1 v_2^*) =\cos\phi\sin\theta \\
	\zps &=& -2 {\rm Im} (v_1 v_2^*)  = \sin\phi\sin\theta. \nonumber 
\end{eqnarray}
A star, *, denotes complex conjugation. We have chosen the polar angle, $\theta$, so that it is measured relative to the x-axis in anticipation of the fact that transits through a cylindrical lens will then lie on arcs of constant $\theta$. Intersections of the SOM with the z-axis ($\theta=\pi/2$, $\phi=\pm\pi/2$) are associated with Laguerre-Gaussian modes, the standard convention~\cite{Padgett1999}.

Position on the SOM is an even less complete description of the state than the original spinors because the spinor phase, $\chi$, has been lost. This can be remedied, though, and the process will point the way towards a means of re-introducing the beam waist information as well. The spinor phase, $\chi$, can be accounted for by generalizing the SOM to a fiber bundle in which a fiber extends out from each point on its surface with $\chi$ measured along such rays. As illustrated in Fig. \ref{Fiber_Space_5}, we can now imagine physical processes that result in two types of motion: transit across the surface of the SOM but also \emph{horizontal lifts} along the fibers~\cite{Kobayashi1996} which track the change in total phase of the beam. 

The generalization from SOM to fiber bundle motivates us to endow each point on the fibers with a 3D SPS in which the beam is assumed to have waists, $w_p$ and $w_a$, aligned with the x and y axes of the original beam, with waist positions of $z_p$ and $z_a$, respectively. Using $z_p=0$ as a datum, the spinors are endowed with a three-element pedigree:
\begin{equation}\label{ket_general}
	v_j = v_j(\phips, \thetaps, \chi ; w_p , w_a , z_a) ,\quad j = 1,2.
\end{equation}
The generalized state space shown in Fig. \ref{Fiber_Space_5} allows state evolution to be represented as a trajectory in which beam waists, waist orientations, and waist positions can all evolve. For a single cylindrical lens oriented with its active axis in the y-direction, $w_1=w_x=w_p$ and $w_2=w_y=w_a$.

\section{State Evolution as a Fraction of Lens Transit}

\subsection{Beam Propagation Through a Cylindrical Lens}

Consider the propagation of a converging 1D $\HG_m$ mode, Eq. \ref{HG1D}, through a thin cylindrical lens of focal length, $f$, positioned at $\zl$ (Fig. \ref{Lens_Focusing_Schematic_2}). The active axis of the lens is aligned with the y-axis, and the mode that emerges will have the same form as Eq. \ref{HG1D} but with a new waist, $w_a$, and waist position, $z_a$. The mode radii along the passive and active axes of the lens have a simple relationship that is sometimes referred to as the Lensmaker's Equation~\cite{Verdeyen1995}:
\begin{equation}
	\frac{1}{f} = \frac{1}{R_p} - \frac{1}{R_a}.
\end{equation}
Using Eq. \ref{radiusgouy}$_2$ and its counterpart for the active axis of the lens, this can be expressed in terms of Rayleigh lengths:
\begin{equation}\label{modematch}
	\frac{1}{f} = \frac{1}{(\zl-z_p)\bigl(1+\frac{\zRp^2}{(\zl-z_p)^2}\bigr)} - \frac{1}{(\zl-z_a)\bigl(1+\frac{\zRa^2}{(\zl-z_a)^2}\bigr)}.
\end{equation}
%
 
Continuity of the beam radius at the lens provides a second equation that also relates the two Rayleigh lengths:
\begin{equation}\label{continuity}
	\frac{\zRp^2 + (\zl-z_p)^2}{\zRp} = \frac{\zRa^2 + (\zl-z_a)^2}{\zRa}.
\end{equation}
%
%
Keeping in mind that we have set $z_p=0$, Eqs. \ref{modematch} and \ref{continuity} can be solved together to give the active-axis Rayleigh length as well as positions in the Supplementary Parameter Space as functions of the lens position, $\zl$:
\begin{eqnarray}\label{SPS}
	z_a &=& f+\zl - \frac{f^2(f-\zl)}{\zRp^2+(f-\zl)^2} \nonumber \\
    \zRa &=& \frac{f^2 \zRp}{\zRp^2 + (f-\zl)^2} \\
	w_a &=& \sqrt{\frac{2 \zRa}{k}}.\nonumber
\end{eqnarray}
%

It is also convenient to define a \emph{natural lens offset}, $-\dnat$, and associated \emph{natural focal length}, $\fnat$, for which $z_a=0$---i.e. a lens position such that the active and passive waists are both at $z=0$:
\begin{eqnarray}\label{f0d0}
	\fnat &=& \zRp \csc(\alpha/2) \biggl( -1+\frac{2}{1+\tan(\alpha/4)}   \biggr) \nonumber \\
	\dnat &=& \zRp (-1 + \frac{2}{1+\tan(\alpha/4)}) .
\end{eqnarray} 
%
%
Here we have borrowed notation from the theory of mode converters~\cite{Beijersbergen1993} by replacing $\dnat/\fnat$ with $\sin(\alpha/2)$. In mode converters, the spacing between lenses is $\dnat$, and $\alpha$ is the difference in Gouy phase between active and passive axes after propagation through the lens, the difference that changes the mode of the beam. This connection is still useful even though our attention is focused on a single lens.

Without making any such restriction on the position of the active waist, the 1D beam that emerges from the lens is
\begin{equation}\label{psiout1D}
	\ket{u^{\rm (out)}} = \ket{m^{w_a,z_a}} e^{i\psi_a(\zl, m)}e^{-i\psi_p(\zl, m)}.
\end{equation}
This implies that a cylindrical lens generates the following change in the total phase of the 1D $HG_m$ mode, a difference of Gouy phases:
\begin{equation}\label{Gouyjump}
	\delg (m,f,\zl) = \psi_a(\zl, m)-\psi_p(\zl, m) .
\end{equation}
%
%
This phase change is an explicit function of lens position, $\zl$, and by virtue of Eqs. \ref{radiusgouy}$_2$ and \ref{SPS}, a function of lens focal length, $f$, as well.

Now extend the approach to quantify the result of sending in a linear combination of 2D HG modes,
\begin{equation}\label{psiin2D}
	\ket{u_{\rm in}} = v_1^{\rm (in)} \ket{1^{w_p,z_p}0^{w_p,z_p}} + v_2^{\rm (in)} \ket{0^{w_p,z_p}1^{w_p,z_p}} ,
\end{equation}
through a cylindrical lens with its active axis aligned in the y-direction. The output beam is then
\begin{equation}\label{psiout2D1}
	\ket{u_{\rm out}} = v_1^{\rm (out)} \ket{1^{w_p,z_p}0^{w_a,z_a}} + v_2^{\rm (out)} \ket{0^{w_p,z_p}1^{w_a,z_a}},
\end{equation}
where the output spinor is
\begin{equation}\label{psiout2D2}
	\underline v^{\rm (out)} = [G] \underline{v}^{\rm (in)}.
\end{equation}
The lens action on the input spinor can be expressed algebraically~\cite{Stoler1981}:
\begin{equation}\label{G}
[G] :=   \begin{pmatrix}
		e^{-i \delg(0,f,\zl)} & 0  \\
		0 & e^{-i \delg(1,f,\zl)}
			\end{pmatrix}   .
\end{equation}
The coordinates, $\underline v^{\rm (out)}$, can then be converted to SOM coordinates using Eq. \ref{PScart}.

It is important to note, though, that this matrix representation does not capture evolution in the SPS.


\subsection{Fractional Lens Transit}
We are now in a position to study beam transit through a cylindrical lens by tracking its trajectory through Supplementary Product Space, along fibers, and across the surface of the Sphere of Modes. This intra-lens description draws on the form of the Thin Lens Approximation~\cite{Saleh2007}, in which the effect of the lens on the beam is taken to be a  multiplicative factor of $ e^{-i \mu y^2}$ with $\mu=z_R/(2 f)$. This implies that passage through N successive lenses would generate a total phase factor of 
\begin{equation}\label{thinlensfrac}
	\prod_{n=1}^N e^{-i\mu_n y^2} \equiv  e^{-i(\sum_{n=1}^N\mu_n) y^2}.
\end{equation}
In this same vein, we may view a single lens as a stack of differentially thin lenses and imagine the beam output after passing through a fraction, $\zeta$, of these as
\begin{equation}\label{zetalens}
	u_{\rm out}(\zeta) = e^{-\frac{y^2}{2 f/\zeta}} u_{\rm in}, \quad \zeta \in [0,1] .
\end{equation}
Jumps in active-axis waist, $w_a$, waist position, $z_a$, fiber phase, $\chi$, and beam structure, $\ket{u_{\rm out}}$, can therefore be transformed to smooth functions of lens transit fraction, $\zeta$, by replacing $f$ with $f/\zeta$ in Eqs. \ref{SPS}, \ref{Gouyjump}, and \ref{psiout2D1}. Parametric plots as a function of $\zeta$ then produce trajectories through the SPS, along fibers and across the surface of the SOM.

\section{Detailed Analysis of Two Different Lens Transits}
Consider the transit of a first-order Gaussian mode through a cylindrical lens with a mode converter phase parameter of $\alpha$, natural focal length, $\fnat$, and general lens position, $\zl$. Assume that the initial fiber phase, $\chi$, is zero. We now have the tools to map out paths across the SOM, along fibers, and through the SPS as the lens transit fraction, $\zeta$, increases from 0 to 1. 

A circular arc will be generated on the SOM for which the polar angle, $\theta$, is constant---a counterclockwise rotation about the x-axis. The change in azimuthal angle, $\Delta\phi$, is obtained from the normalized inner product of the Stokes coordinates just before and just after the lens. Eqs. \ref{Jones2}, \ref{PScart}, and \ref{psiout2D2} then imply that
\begin{eqnarray}\label{arc}
	\Delta\phi &=&\cos^{-1}\biggl[\frac{\underline v^{\rm (out)}.\underline v^{\rm (in)}}{|\underline v^{\rm (out)}||\underline v^{\rm (in)}|} \biggr] \nonumber \\
	&=& \tan^{-1}(\zl) + \tan^{-1}\biggl(\frac{1+\zl(\zl-f)}{f}\biggr).
\end{eqnarray}
The arc length is a function of the lens position, $\zl$. Placements further to the left of the passive waist position ($z=0$) will decrease the arc length; likewise the arc length increases as the lens is moved further to the right. Eq. \ref{arc} implies that the maximum arc length for any lens is $\pi$.

\begin{figure}[t]
	\begin{center}
		\includegraphics[width=0.8\linewidth]{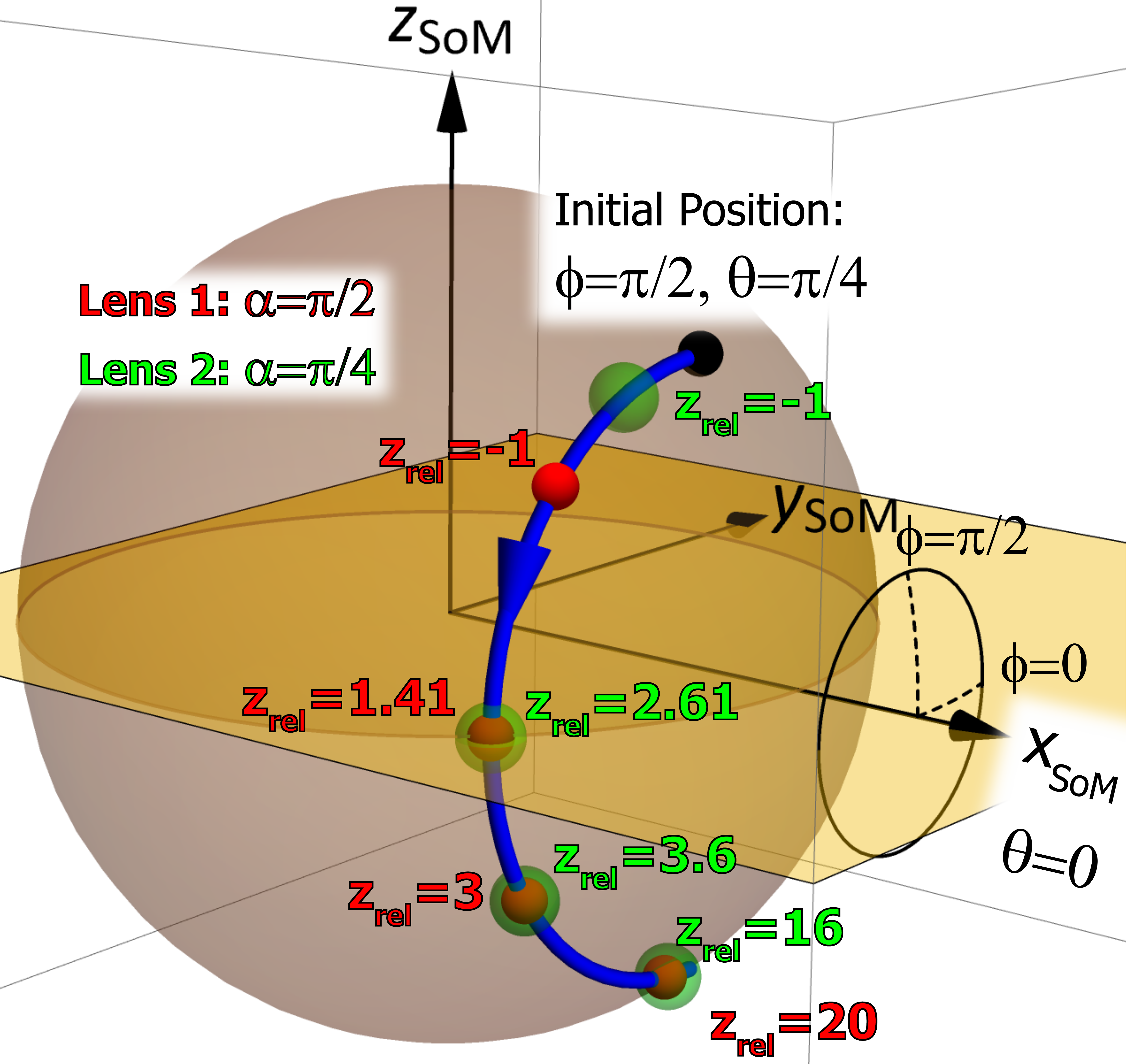}
	\end{center}
	\caption{ \emph{Sphere of Modes Trajectories for Lenses 1 and 2}. A Gaussian beam, Eq. \ref{Jones1}, with spinor state Eq. \ref{Jones2} and SOM angles of $\phi=\pi/2, \, \theta=\pi/4$ (black sphere, $\zeta=0$), transits two individual cylindrical lenses in separate experiments. Their focal lengths are described by Eq. \ref{f0d0}$_1$ with $\alpha_1=\pi/2$ and $\alpha_2=\pi/4$. As a reminder, $\alpha$ is the difference in Gouy phase between active and passive axes after propagation through the lens, the difference that changes the mode of the beam. The spinor evolves according to Eq. \ref{psiout2D2}, with $f$ replaced by $f/\zeta$, generating the blue curve shown. The length of this trajectory depends on lens position, $\zl=\zrel \dnat$, with $\dnat$ given by Eq. \ref{f0d0}$_2$ for each lens. In all cases, the endpoint of an arc is associated with $\zeta=1$.}
	\label{PS_Traj_Both_Cases}
\end{figure}

For the sake of making a concrete comparison, two sets of single-lens transits are considered in Fig. \ref{PS_Traj_Both_Cases}. Each set is for a particular lens so they are labeled as Lens 1 $\Leftrightarrow \alpha=\pi/2$ and Lens 2 $\Leftrightarrow \alpha=\pi/4$. For each lens, the SOM trajectory is calculated for a range of lens positions. For both sets, the pre-lens mode of the beam ($\zeta=0$) is associated with position $\{ \phi=\pi/2, \theta=\pi/4\}$, indicated by the small black sphere in Fig. \ref{PS_Traj_Both_Cases}. All trajectories, for both lenses, are shown to lie on the small-circle, for which $\theta=\pi/4$. As a given lens is moved further from the source, the arc length increases, and the trajectory angle subtended approaches the maximum value of $\pi$. Note, in particular, that each lens can be positioned so as to generate an arc angle of $\pi/2$, but that the lenses must be placed in distinctly different positions for this to occur: $\zl=0.59 \zRp\,(\zrel=1.41)$ and $\zl=1.77 \zRp\,(\zrel=2.61)$, respectively.

%
%
%

\begin{figure}[t]
	\begin{center}
		\includegraphics[width=0.8\linewidth]{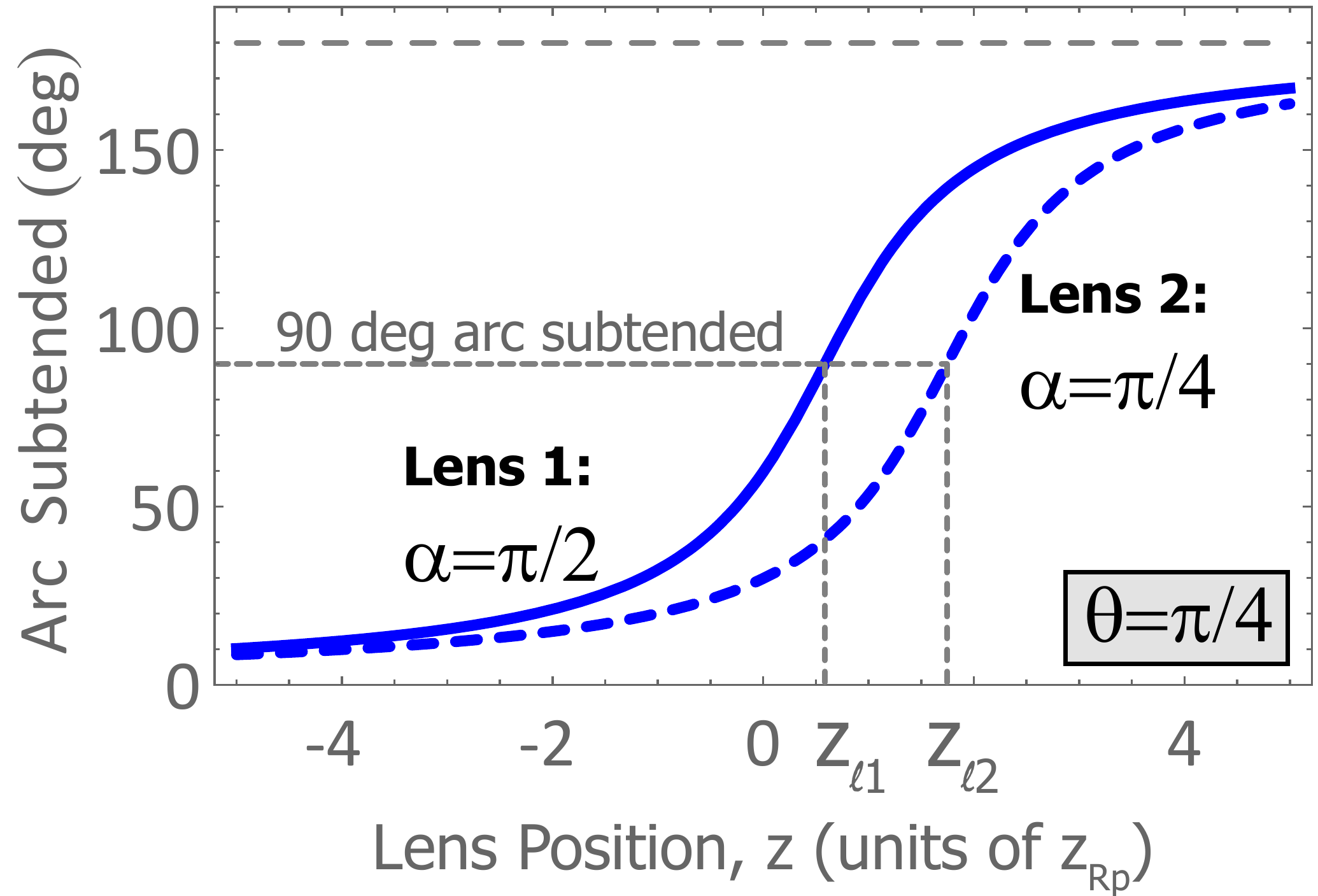}	
	\end{center}
	\caption{ \emph{Arc Length of Sphere of Modes Trajectories}. As shown in Fig. \ref{PS_Traj_Both_Cases}, the complete trajectory arc length is a function of lens position, $\zl$. The total angle subtended is given by Eq. \ref{arc} and is plotted here for Lens 1 and Lens 2. The dashed line is at an arc angle of $90^{\circ}$, and the associated lens positions are $\zla = 0.59 \zRp$ and $\zlb = 1.77 \zRp$. These correspond to $\zrela = 1.41$ and $\zrelb = 2.61$.}
	\label{Arc_Length_v_Lens_Pos}
\end{figure}

The blue arc in Fig. \ref{PS_Traj_Both_Cases} shows the trajectory traced out on the SOM, in association with each lens, as a function of lens transit fraction. The arc endpoints ($\zeta=1$) shown with red (darker, smaller) spheres are for Lens 1 ($\alpha=\pi/2$) while the green (larger, lighter) spheres are for lens 2 ($\alpha=\pi/4$). These relative position labels, $\zrel$ are given in terms of the respective natural lens displacements, $\dnat$. Fig. \ref{Arc_Length_v_Lens_Pos} shows how these arc lengths change as a function of lens position, now in units of the passive Rayleigh length, $\zRp$.

%
%
%

\begin{figure}[t]
	\begin{center}
		\includegraphics[width=0.8\linewidth]{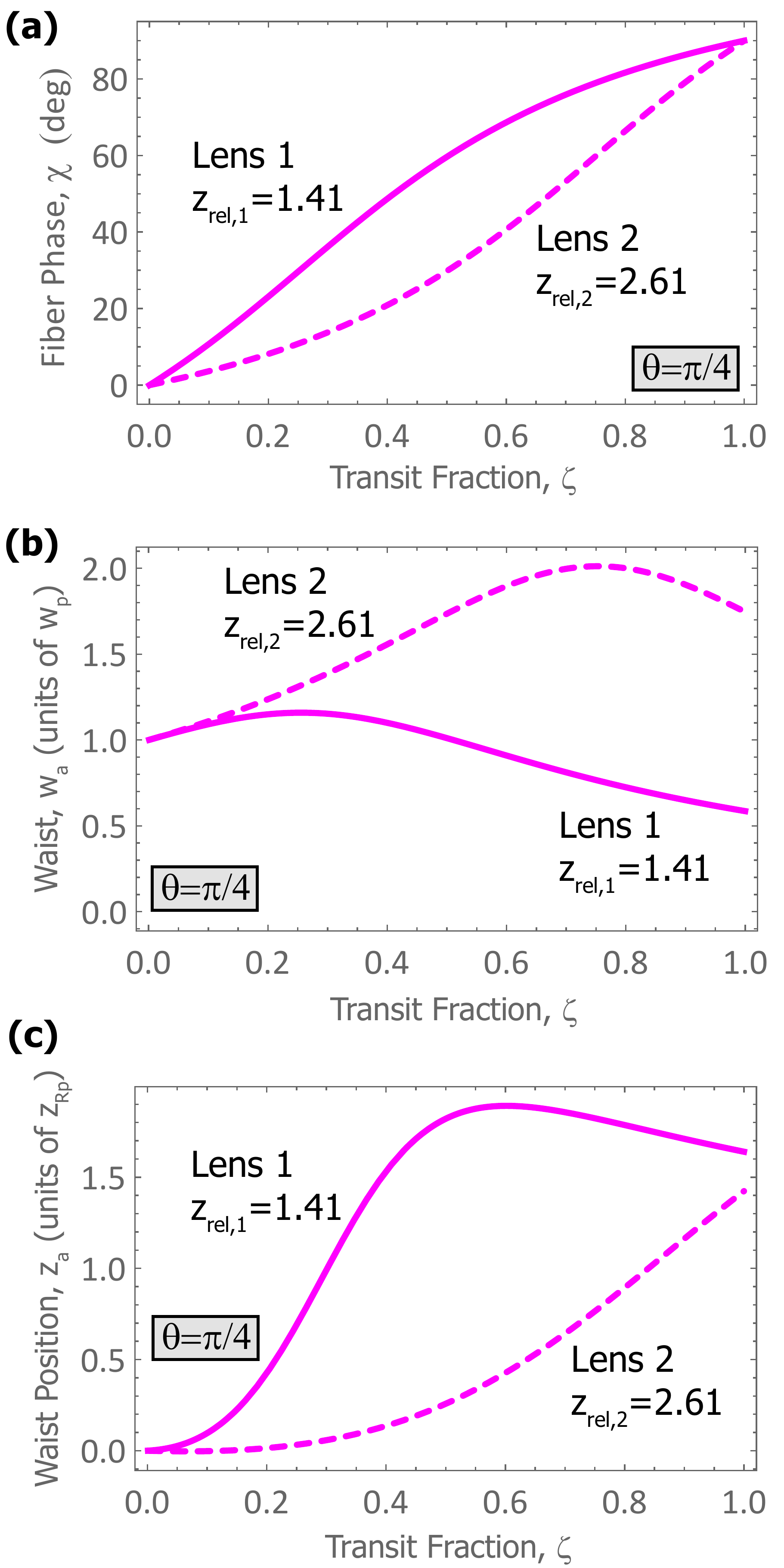}
	\end{center}
	\caption{ \emph{Trajectory Through the Supplementary Product Space}. Figs. \ref{PS_Traj_Both_Cases} and \ref{Arc_Length_v_Lens_Pos} show the Sphere of Modes trajectories associated with beam transit through two lenses, and here the corresponding trajectories along fibers and through the Supplementary Product Space are plotted using Eqs. \ref{SPS}$_{2,3}$ and \ref{Phitot}. The lens positions, $\zrela = 1.41$ and $\zrelb = 2.61$, result in the same SOM trajectories with an arc angle of $90^{\circ}$, a convenient reference point for comparing the geometric phase accumulated in each case. Here $\zl=\zrel \dnat$ with $\dnat$ given by Eq. \ref{f0d0}$_2$.}
	\label{Suppl_Space_Transit_Frac}
\end{figure}

A trajectory along fibers and through the SPS accompanies the SOM trajectory of Fig. \ref{PS_Traj_Both_Cases}.  The evolving fiber phase, $\chi$, is obtained using Eqs. \ref{Jones2}, \ref{PScart}, and \ref{psiout2D2}:
\begin{equation}\label{Phitot}
	\chi(\zeta) = \tan^{-1}\zl + \tan^{-1}\biggl[\frac{-f \zl + \zeta + \zl^2\zeta}{f}\biggr].
\end{equation}
The resulting fiber phase is plotted in Fig. \ref{Suppl_Space_Transit_Frac}(a) for the same parameters as Figs. \ref{PS_Traj_Both_Cases} and \ref{Arc_Length_v_Lens_Pos}.

%
%
%

For each lens, a trajectory through the SPS can be generated using Eqs. \ref{SPS}$_{2,3}$. As noted previously, the focal length, $f$, must be replaced by $f/\zeta$ in these equations. The resulting progress through the SPS is plotted in panels (b) and (c)  of Fig. \ref{Suppl_Space_Transit_Frac}. 

\begin{figure}[t]
	\begin{center}
		\includegraphics[width=1\linewidth]{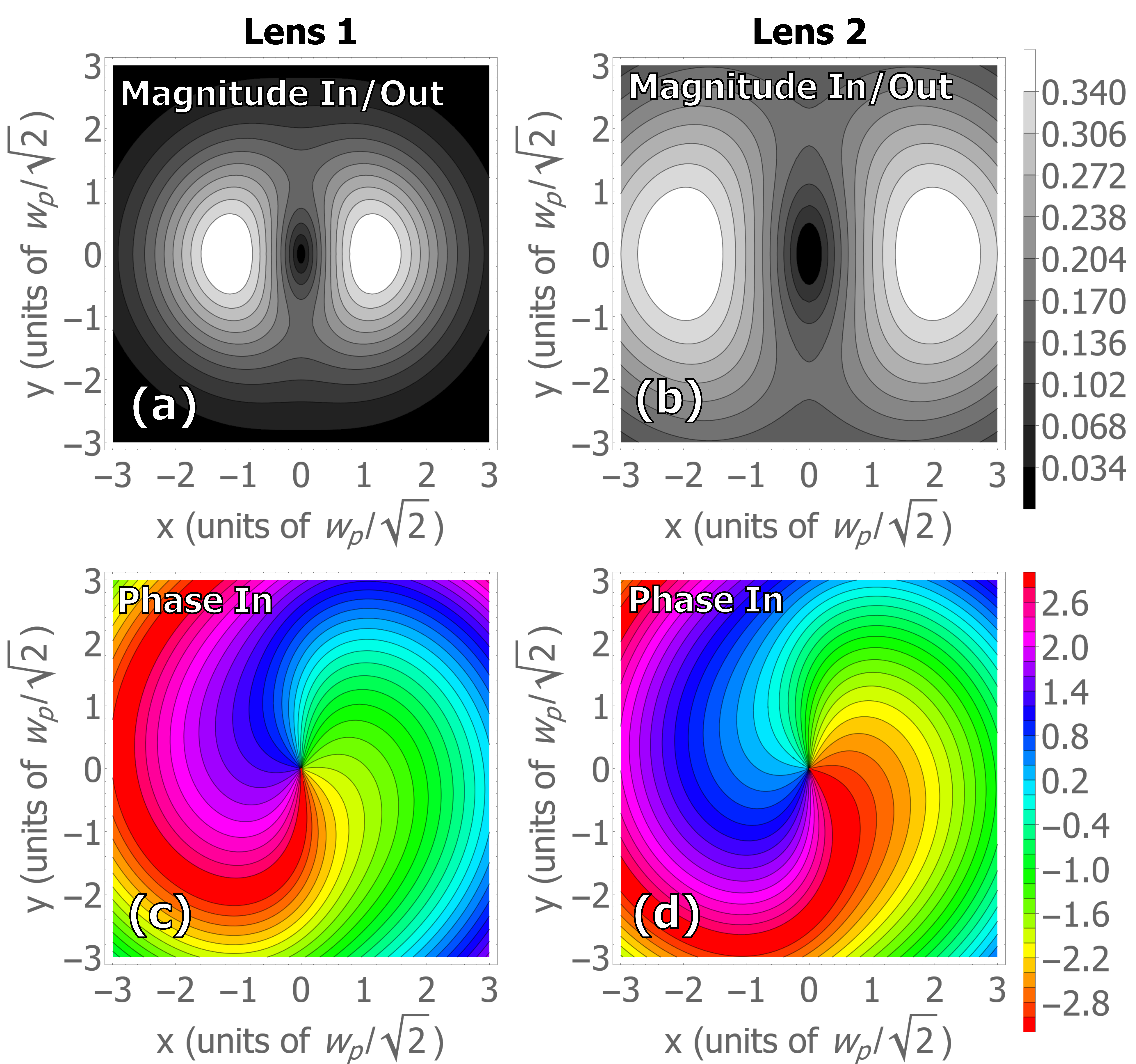}
	\end{center}
	\caption{ \emph{Beam Cross-Sections Before Each Lens}. The information used to plot the trajectories on the Sphere of Modes (Figs. \ref{PS_Traj_Both_Cases} and \ref{Arc_Length_v_Lens_Pos}), motion along the fibers (Fig. \ref{Suppl_Space_Transit_Frac}(a)), and trajectories through the Supplementary Product Space (Fig. \ref{Suppl_Space_Transit_Frac}(b,c)), can all be combined in Eq. \ref{Jones1} to produce beam cross-sections that depend on the transit fraction, $\zeta$. The beam magnitude has a cross-section that does not depend on transit fraction, so a single magnitude cross-section is plotted for each lens in panels (a) and (b). Panels (c) and (d) give the cross-sectional phase of the beam entering the lens. These look similar, but they are clearly not the same. This is because the lens positions are different, $\zrela = 1.41$ and $\zrelb = 2.61$, the values used to produce Fig. \ref{Suppl_Space_Transit_Frac}.}
	\label{Beam_Structure_In}
\end{figure}

\begin{figure}[t]
	\begin{center}
		\includegraphics[width=1\linewidth]{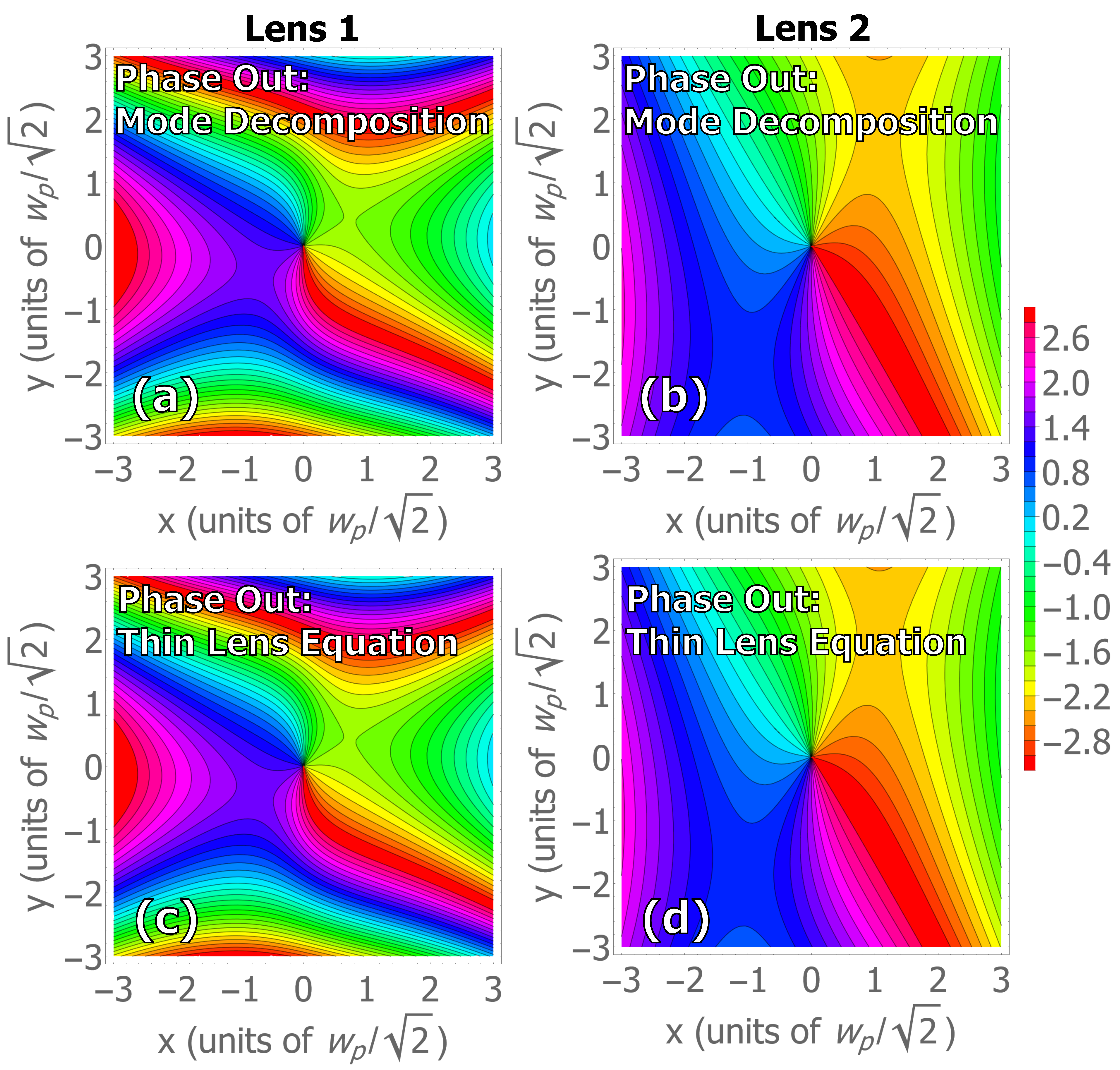}
	\end{center}
	\caption{ \emph{Beam Cross-Sections After Each Lens}. The information used to plot the trajectories on the Sphere of Modes (Figs. \ref{PS_Traj_Both_Cases} and \ref{Arc_Length_v_Lens_Pos}), motion along the fibers (Fig. \ref{Suppl_Space_Transit_Frac}(a)), and trajectories through the Supplementary Product Space (Fig. \ref{Suppl_Space_Transit_Frac}(b,c)), can all be combined in Eq. \ref{Jones1} to produce beam cross-sections that depend on the transit fraction, $\zeta$. Panels (a) and (b) are the corresponding cross-sectional phases for the beams exiting the lenses, and these are clearly different. Panels (c) and (d) are verification plots in which the Thin Lens Equation, Eq. \ref{thinlens}, is used to plot the phase cross-sections directly. These plots are identical to mode-based plots of panels (a) and (b).}
	\label{Beam_Structure_Out}
\end{figure}

The combination of spinor coefficients, $v_1$ and $v_2$, along with fiber phase, $\chi$, active beam waist, $w_a$, and active waist position, $z_a$, is sufficient to generate beam cross-sections before and after each lens. The results shown in Figs. \ref{Beam_Structure_In} and \ref{Beam_Structure_Out} are based on the same information as was used to produce Figs. \ref{PS_Traj_Both_Cases}, \ref{Arc_Length_v_Lens_Pos}, and \ref{Suppl_Space_Transit_Frac}, but it is useful to see that identical start and end points of the SOM trajectory can be associated with such distinctly different beam cross-sections for lenses 1 and 2. 

As an important verification step, the beam that emerges from each lens can also be described using the Thin Lens Equation~\cite{Saleh2007}:
\begin{equation}\label{thinlens}
	u_{\rm out} = u_{\rm in}e^{-i y^2/(2f)}.
\end{equation}
Cross-section plots using this equation are plotted in panels (c) and (d) of Fig. \ref{Beam_Structure_Out}, and it is clear that they are identical to the results shown in panels (e) and (f), respectively. Of course, the form of Eq. \ref{thinlens} does not allow any conclusions to be obtained concerning the evolution of basis modes as a function of lens transit, nor does it provide any information about evolution along fibers and through the SPS.

%
%
%

%
%
%

\section{Geometric Phase}
\begin{figure}[t]
	\begin{center}
		\includegraphics[width=0.6\linewidth]{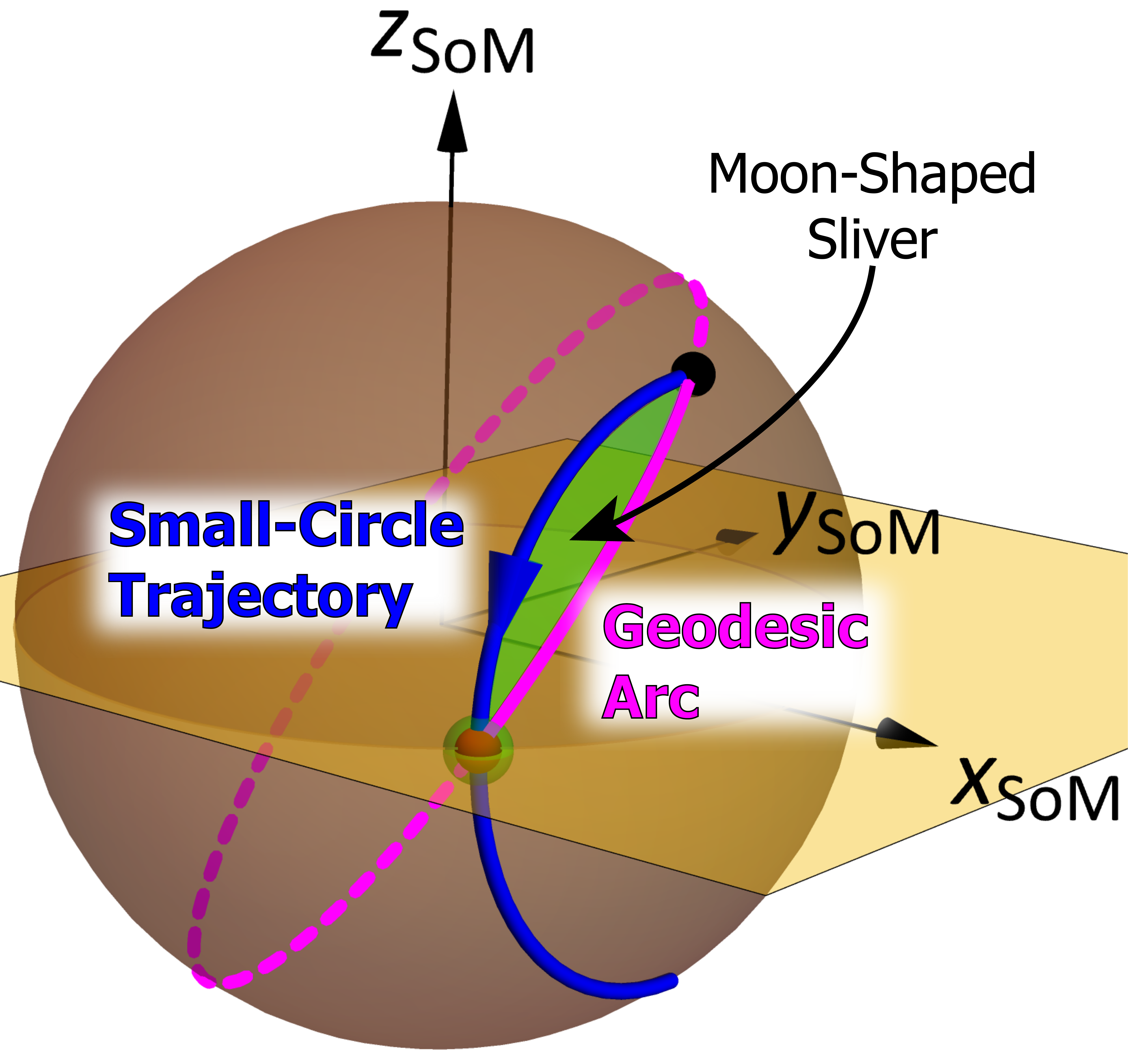}
	\end{center}
	\caption{ \emph{Intersection of Small-Circle Arc and Geodesic}. The geometric phase accumulated by beam transit through a lens is quantified using Eq. \ref{PhiPanch}. This requires that the solid angle subtended by the moon-shaped sliver, shown here, between geodesic and small-circle arcs be evaluated. Eq. \ref{moon} gives its solid angle.}
	\label{Geodesic_Arc}
\end{figure}

It is now a well-told story that an experimentally measurable geometric phase~\cite{Berry1985, Simon1983} can be identified with  non-adiabatic dynamics~\cite{Aharonov1987} and does not require that trajectories on the SOM be closed~\cite{Pancharatnam1956, Samuel1988, Mukunda1993}. A gauge-invariant geometric phase associated with the \emph{non-geodesic} arcs, such as the small-circle trajectories of Fig. \ref{PS_Traj_Both_Cases}, can be obtained by closing the arc with a geodesic, as shown in Fig. \ref{Geodesic_Arc} and evaluating the difference between the total and dynamic phases associated with the complete circuit:
\begin{equation}
\Phi_{\rm geom}=\Phi_{\rm tot} - \Phi_{\rm dyn} .
\end{equation}

The geodesic arc is the Stokes manifestation of a projection operation carried out on the beam that emerges from our lens:
\begin{equation}\label{projection_1}
	\ket{u_{\rm proj}}= \hat P\ket{u_{\rm out}}.
\end{equation}
where $\ket{u_{\rm out}}$ is given in Eq. \ref{psiout2D1}. The projection operator, $\hat P = \ket{u_{\rm proj}} \bra{u_{\rm proj}}$, is defined in terms of a vector with the same SOM position as the input to the lens, but it has the SPS coordinates of the beam coming out of the lens: 
\begin{equation}\label{projection_2}
\ket{u_{\rm proj}} =	v_1^{\rm (in)} \ket{1^{w_p,z_p}0^{w_a,z_a}} + v_2^{\rm (in)} \ket{0^{w_p,z_p}1^{w_a,z_a}}.
\end{equation}
The projection operation therefore produces a final state vector that results in a closed SOM circuit while not returning the beam to its original state:
\begin{equation}\label{projection_3}
	\ket{u_{\rm final}} =\underline v^{\rm (in)*} \cdot \underline v^{\rm (out)}   \ket{u_{\rm proj}}.
\end{equation}
%

%
%
%

\begin{figure}[t]
	\begin{center}
		\includegraphics[width=0.8\linewidth]{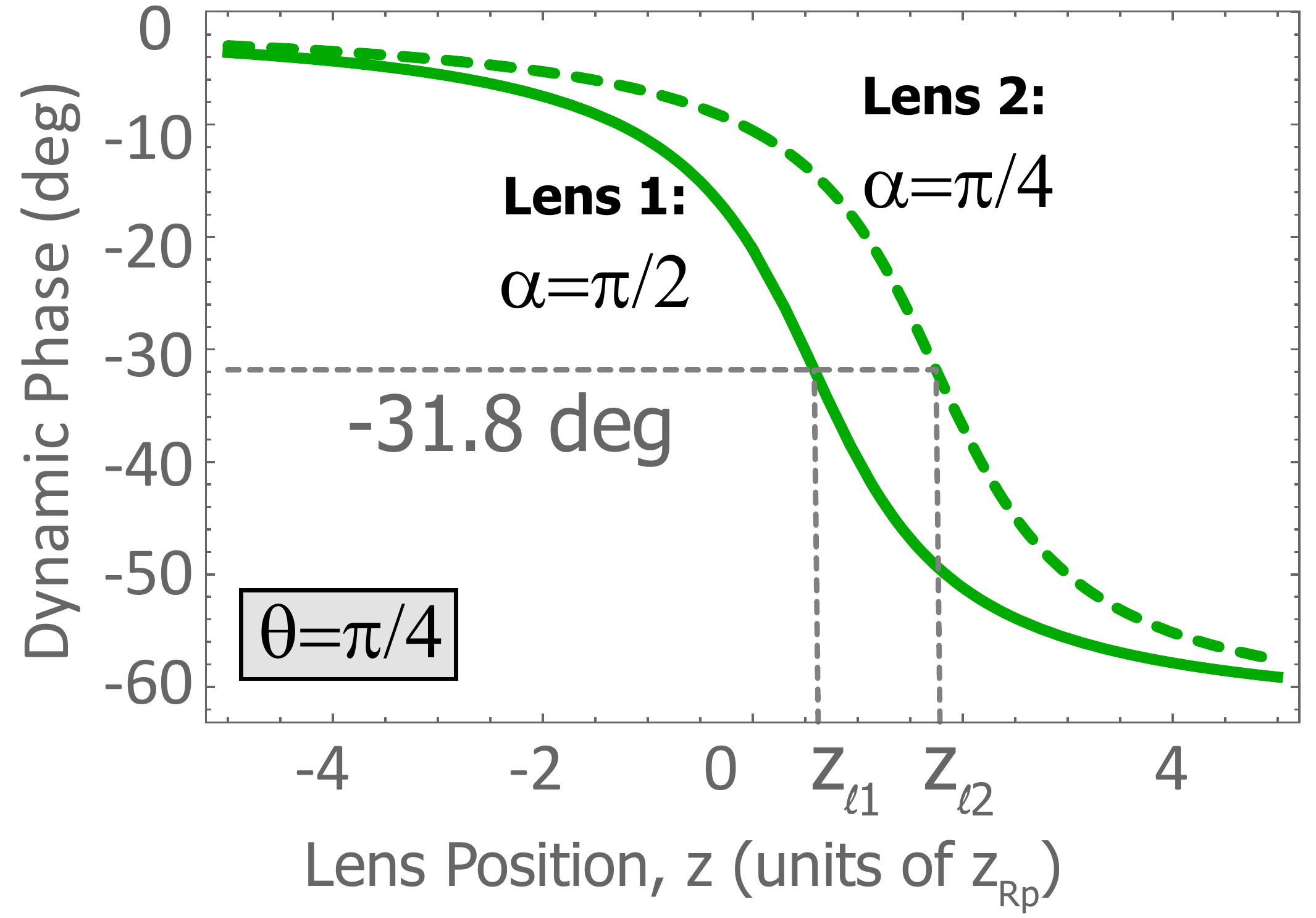}
	\end{center}
	\caption{ \emph{Dynamic Phase Accumulation as a Function of Lens Position}. Eqs. \ref{phi_dyn_2} and \ref{arc} are combined to generate the dynamic phase accumulated by beam transits through Lens 1 and Lens 2, followed by a projection operation, as plotted on the SOM in Fig. \ref{PS_Traj_Both_Cases}. This phase depends on lens position, $\zl$. For lens positions of $\zla = 0.59 \zRp$ and $\zlb = 1.77 \zRp$, the SOM trajectories are the same for each lens ($90^{\circ}$ arc angles) and the dynamic phase is evaluated to be the same, $-31.8^\circ$, for the two lenses.}
	\label{Dynamic_Phase_v_Lens_Position}
\end{figure}

In the current setting, where dynamic phase~\cite{Aharonov1987} accumulates as a function of azimuthal angle subtended, $\phi$, the phase can be defined with respect to the Jones vector as
\begin{equation}
	\Phi_{\rm dyn}(\phi,\theta) = -i \int_{\phii}^{\phii+\Delta\phi} d\phi' \underline v^*(\phi',\theta)   \cdot \partial_{\phi'}\underline v(\phi',\theta).
	\label{phi_dyn_1}
\end{equation}
Here the fiber phase has been removed. Using Eq. \ref{Jones2}, this implies that
\begin{equation}
	\Phi_{\rm dyn}(\Delta\phi,\theta) = -\frac{\Delta\phi}{2}\cos\theta ,
	\label{phi_dyn_2}
\end{equation}
a useful result already in the literature for polarization~\cite{Bhandari1989}. Within the thin-lens approximation, this dynamic phase accumulates because of a change of basis and not in association with any physical propagation. For transit through a single lens, Eq. \ref{arc} can be used to express this as a function of lens position, $\zl$, and the result is plotted in Figure \ref{Dynamic_Phase_v_Lens_Position}.
%
%
%
%

The same expression can also be used to show that the dynamic phase accumulated over any portion of a geodesic is zero since the associated polar angle, $\theta$, is equal to $\pi/2$ in Eq. ~\ref{phi_dyn_2}. This has the geometric interpretation that the requisite closure of a geodesic arc simply re-traces the arc. Eq. \ref{phi_dyn_2} therefore gives the total dynamic phase accumulated over the closed circuit shown in Fig. \ref{Geodesic_Arc}.

\begin{figure}[t]
	\begin{center}
		\includegraphics[width=0.8\linewidth]{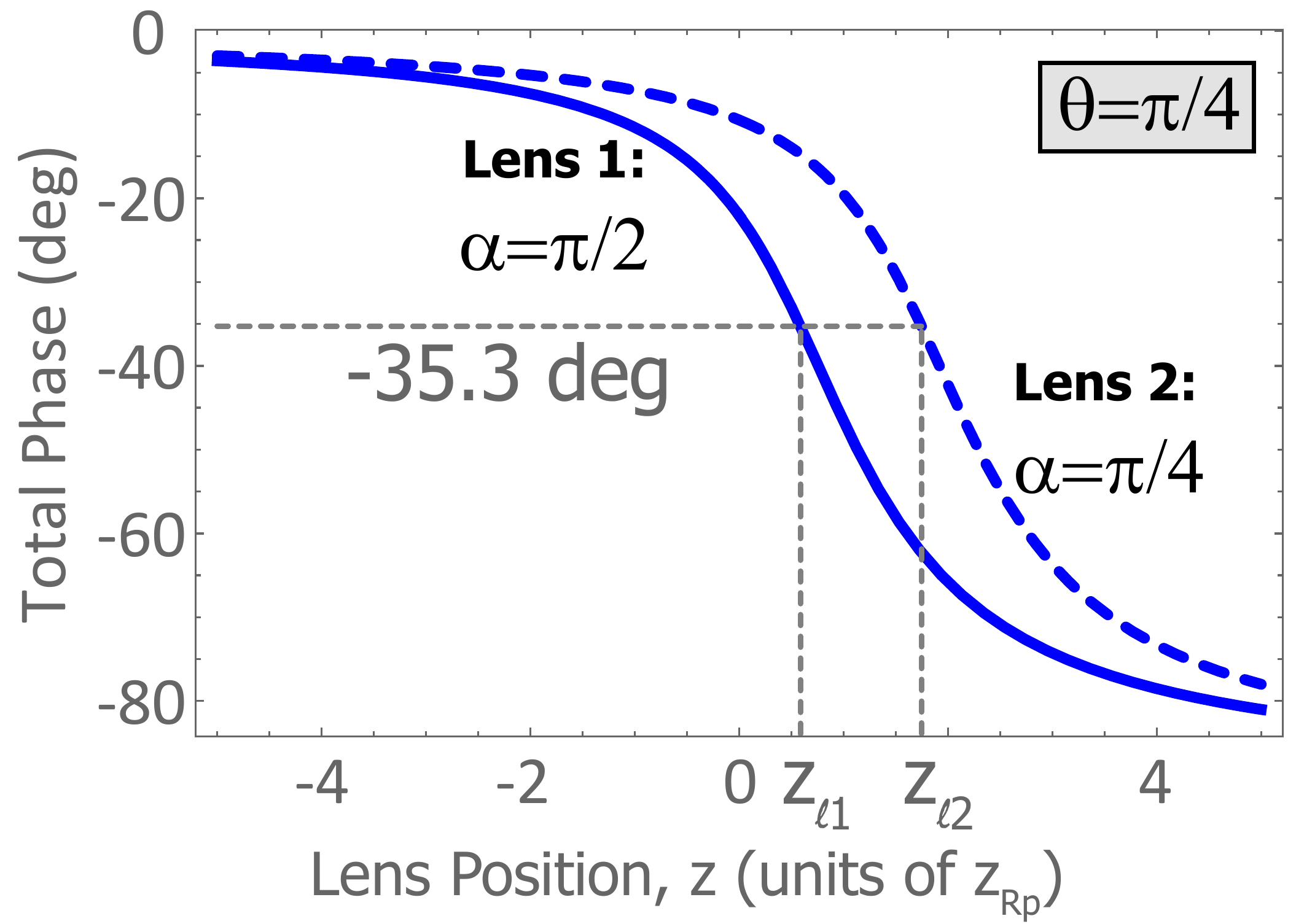}
	\end{center}
	\caption{ \emph{Total Phase Accumulation as a Function of Lens Position}. Eqs. \ref{phi_tot_4} and \ref{arc} are combined to generate the total phase accumulated by beam transits through Lens 1 and Lens 2, followed by a projection operation, as plotted on the SOM in Fig. \ref{PS_Traj_Both_Cases}. This phase depends on lens position, $\zl$. For lens positions of $\zla = 0.59 \zRp$ and $\zlb = 1.77 \zRp$, the SOM trajectories are the same for each lens ($90^{\circ}$ arc angles) and the total phase is evaluated to be the same, $-35.3^\circ$, for the two lenses.}
	\label{Total_Phase_v_Lens_Position}
\end{figure}

Total phase, $\Phi_{\rm tot}$, is typically expressed~\cite{Aharonov1987} so that the state at the end of a closed circuit is the same as that at the beginning except for a phase factor of $e^{i \Phi_{\rm tot}}$. With a SPS, though, the two state need only have the same SOM projection, so that the total phase can be generalized by expressing it in terms of Jones vectors of the state going into the lens, $\ket{u_{\rm in}}$, and the state produced by the projection operation, $\ket{u_{\rm proj}}$:
\begin{equation}
	\Phi_{\rm tot} = \arg(\underline v^*_{\rm in} \cdot \underline v_{\rm out})
	\label{phi_tot_2} .
\end{equation}
As with the dynamic phase, the evolving fiber phase has been removed. When applied to beam transit through a single lens, this is
\begin{equation}
\Phi_{\rm tot}^{\rm lens}= -\tan^{-1}\bigl(\cos(\Delta\phi/2), \cos(\theta) \sin(\Delta\phi/2) \bigr).
	\label{phi_tot_4}
\end{equation}
Since the geodesic arc represents a projection operation, though, it does not contribute any total phase accumulation. The total phase of Eq. \ref{phi_tot_4} therefore applies to the entire closed circuit shown in Fig. \ref{Geodesic_Arc}. As with the dynamic phase, Eq. \ref{arc} can be used to plot this as a function of lens position, carried out in Fig. \ref{Total_Phase_v_Lens_Position} for Lens 1 and Lens 2.
%
%
%
%

Geometric phase is defined as the difference between the total and dynamic phase~\cite{Aharonov1987}, and it can be immediately calculated using Eqs. \ref{phi_dyn_2} and \ref{phi_tot_4}.

As a check on this approach for dissecting geometric phase, the reliance on differential parallelism~\cite{Simon1983} can be replaced with Pancharatnam's method~\cite{Pancharatnam1956} of quantifying the degree of parallelism between two arbitrarily separated states. The geometric phase derived in this way has a nice geometric interpretation; it is negative one-half the solid angle, $\Omega_{\rm \scriptstyle moon}$, subtended by the moon-shaped sliver of Fig. \ref{Geodesic_Arc}:
\begin{equation}\label{PhiPanch}
	\Phi_{\rm geom, lens}^{\rm Panch} = -\frac{\Omega_{\rm \scriptstyle moon}}{2}.
\end{equation}
Here
\begin{eqnarray}\label{moon}
	\Omega_{\rm \scriptstyle moon} &=& \Delta\phi(1-\cos\theta) \nonumber \\
	&-2&\, \cot^{-1}\biggl[ \bigl( \cot\Delta\phi + \frac{(\cot\theta + \csc\theta)^2}{\sin\Delta\phi} \bigr) \biggr] .
\end{eqnarray}
%
%
Although equivalent to using Pancharatnam's \emph{In-Phase Rule}, this expression was obtained by simply calculating the intersection of a curvilinear tetrahedron~\cite{Eriksson1990} and a polar cap of constant latitude. As usual, $\Delta\phi$ and $\theta$ refer to the azimuthal span and polar angle with respect to the x-axis, respectively.

\begin{figure}[t]
	\begin{center}
		\includegraphics[width=0.8\linewidth]{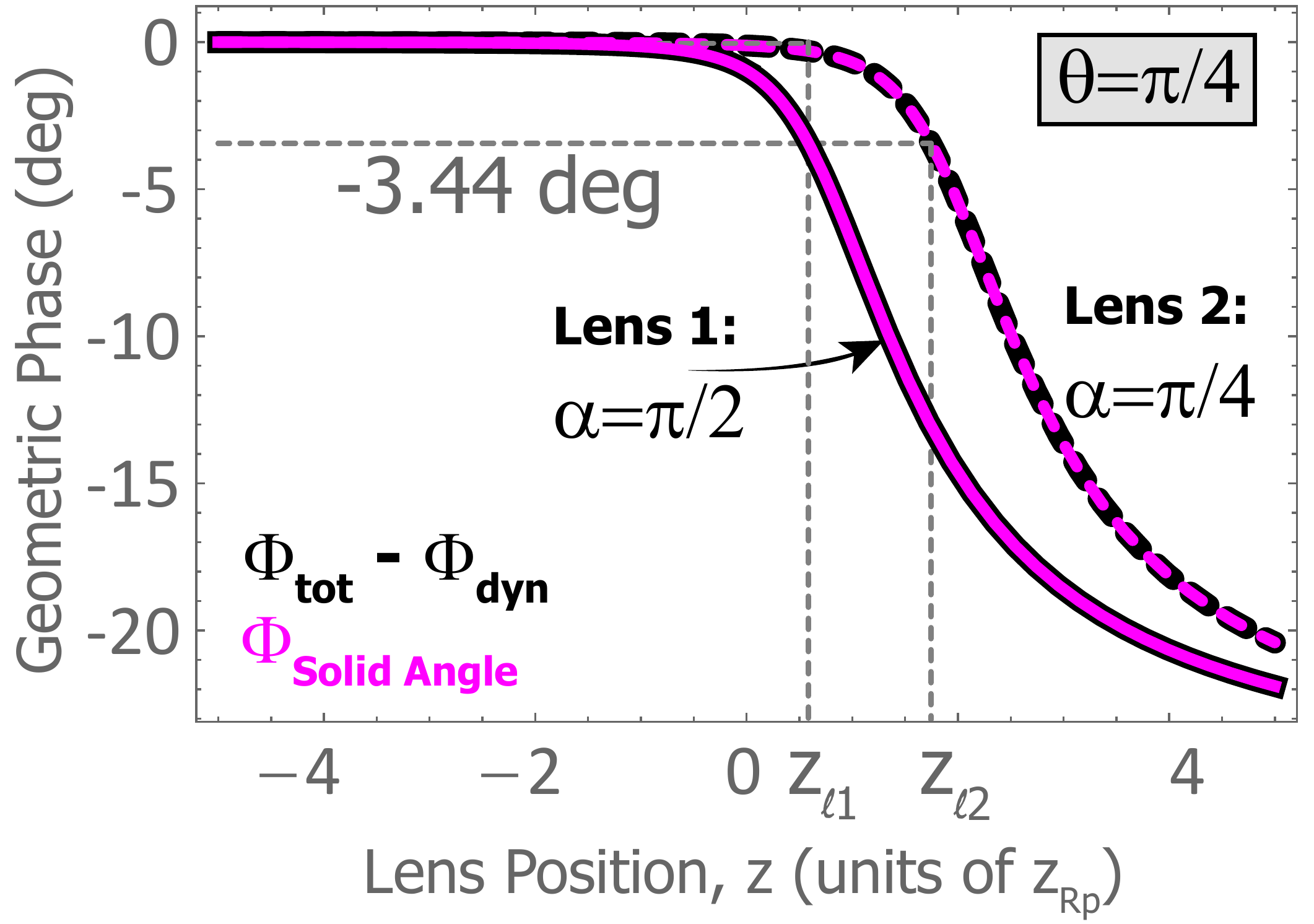}
	\end{center}
	\caption{ \emph{Geometric Phase Accumulation as a Function of Lens Position}. Eqs. Eqs. \ref{phi_dyn_2} and \ref{phi_tot_4} or, equivalently, Eq. \ref{PhiPanch}, can be used to generate the geometric phase accumulated by beam transits through lenses 1 and 2, as plotted on the SOM in Fig. \ref{PS_Traj_Both_Cases}. This phase depends on lens position, $\zl$. For lens positions of $\zla = 0.59 \zRp$ and $\zlb = 1.77 \zRp$, the SOM trajectories are the same for each lens ($90^{\circ}$ arc angles) and the geometric phase is evaluated to be the same for the two lenses.}
	\label{Geometric_Phase_v_Lens_Position}
\end{figure}

Now we can compare our derivation of geometric phase, using Eqs. \ref{phi_dyn_2} and \ref{phi_tot_4}, to the Pancharatnam result of Eq. \ref{PhiPanch}. This is carried out for the two different lens transits shown in Fig. \ref{PS_Traj_Both_Cases} to produce the comparisons of geometric phase accumulation plotted in Fig. \ref{Geometric_Phase_v_Lens_Position}. The results match perfectly. Note that the two lenses produce the same geometric phase, $3.44^{\circ}$, if Lens 1 is positioned at $\zla$ and Lens 2 is positioned at $\zlb$. This is expected since the geometric phase does not depend on the fiber phase, $\chi$, or position within the SPS. 
%
%
%
%

\begin{figure}[t]
	\begin{center}
		\includegraphics[width=0.8\linewidth]{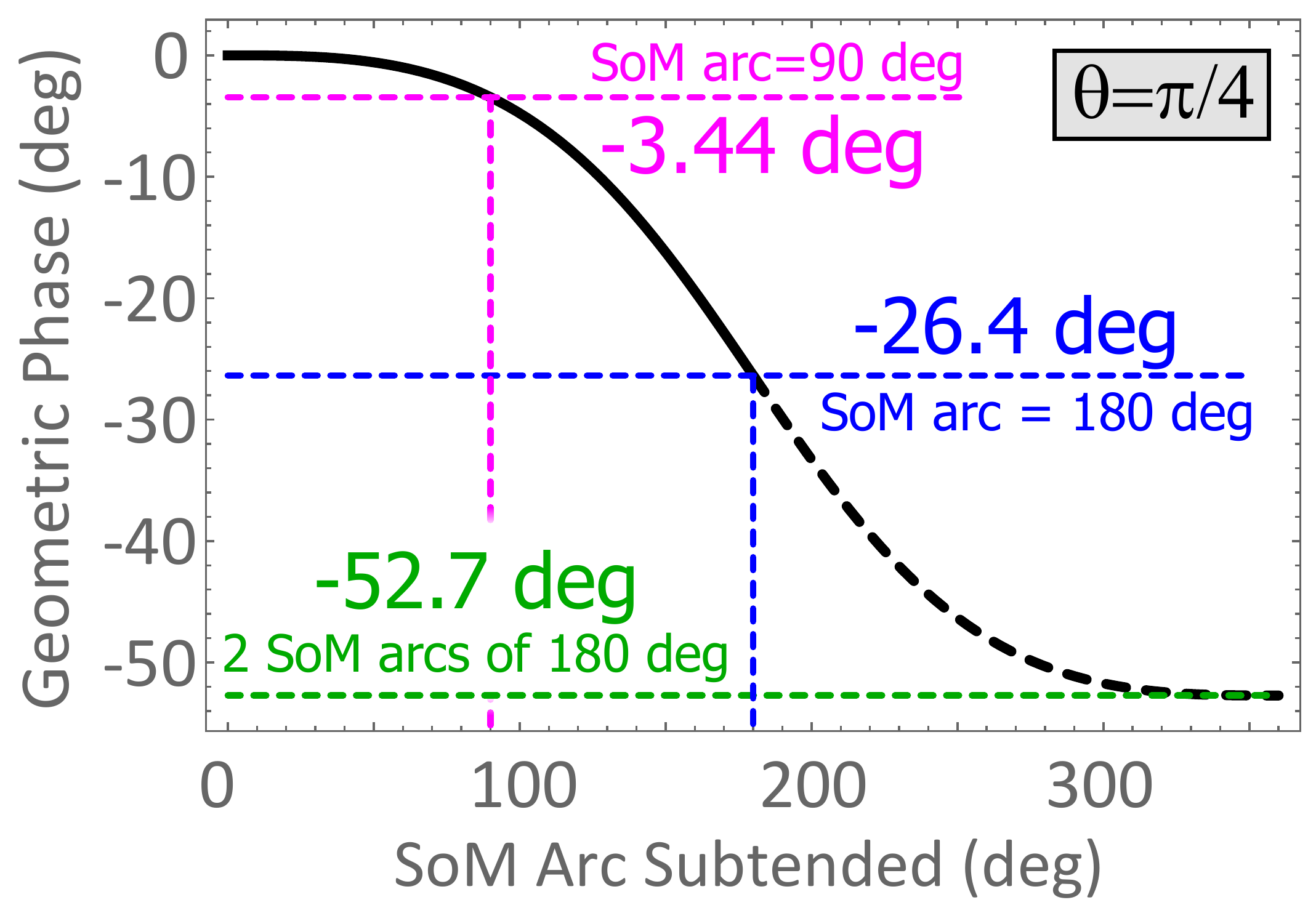}
	\end{center}
	\caption{ \emph{Geometric Phase as a Function of Arc Subtended}. The geometric phase accumulated by beam transit through a lens, Eq. \ref{PhiPanch}, can be used to plot the geometric phase accumulated as a function of the SOM arc subtended. The maximum magnitude for a single lens, $-26.4^{\circ}$, occurs for an arc angle of $180^{\circ}$, while a complete small circle would generate twice as much geometric phase, $-52.7^{\circ}$. This is equal to negative one-half the solid angle enclosed by a polar cap of polar angle $\pi/4$, as given by Eq. \ref{cap}.}
	\label{Geometric_Phase_v_Arc_Length}
\end{figure}

The nonlinear relationship between SOM arc subtended and geometric phase is shown in Fig. \ref{Geometric_Phase_v_Arc_Length}, where it is shown that the maximum magnitude of geometric phase that can be accumulated for Lens 1 or Lens 2 is $-26.36^{\circ}$. However, Eq. \ref{PhiPanch} can be used to calculate the geometric phase of an arbitrary small-circle arc beyond the maximum arc angle of $\pi$ that can be generated with a single lens. This is relevant to a series of mode converters for instance. As shown in the figure, Eq. \ref{PhiPanch} has the expected asymptote of negative one-half the solid angle of a spherical cap centered on the x-axis:
\begin{equation}\label{cap}
	\Omega_{\rm cap} = 4\pi\sin^2(\theta/2),
\end{equation}
where $\theta$ is the polar angle as measured from the x-axis. 
%
%
%
%

Eqs. \ref{PhiPanch} and \ref{moon} can be readily applied to quantify the geometric phase accumulated across any cylindrical lens. To calculate the geometric phase of one or more mode converters, for instance, we only need to know the input and output spinor for each lens. The geometric phase of the converter is then the sum of that of its components. In the same way, a closed circuit composed of a series of mode converters can be calculated in this way as well. Of course, it is usually easier to identify the geometric phase as being negative one-half the solid angle subtended by the closed circuit, but this comes at the cost of losing all information about modes. 

\section{Lens Trajectories are Circular Arcs}
In the examples we considered, lens trajectories on the SOM were found to lie on circles, and this will be true for any lens transit. Consider an abstracted setting in which the input beam to a lens is described by the following 2-parameter spinor:
\begin{equation}\label{circ}
	\underline{v}^{in} = \begin{pmatrix}
		\cos(\mu/2)  \\
		\sin(\mu/2)e^{-i \nu} 
	\end{pmatrix}.
\end{equation}
The effect of the lens on this state is to change the phase along one axis. Rather than relating it to lens parameters, the change can be parametrized as a linear function of lens transit fraction, $\zeta$:  
\begin{equation}\label{circG}
	[G] := \begin{pmatrix}
		1 & 0  \\
		0 & e^{-i \zeta} 
	\end{pmatrix}.
\end{equation}
Within this abstract setting, we can also account for a misalignment of the lens axes with respect to the principal axes of the beam by simply considering a different basis to describe a beam; specifically, we can always choose a basis that is aligned with the lens axes. The spinor, as a function of lens transit fraction, $\zeta$, is therefore just $\underline{v}(\zeta) = [G] \underline{v}^{\rm (in)}$. Eq. \ref{PScart} then gives the following SOM coordinates for the trajectory:
\begin{eqnarray}\label{circPS}
	x &=& \cos(\mu) \nonumber \\
	y &=& \cos(\zeta + \nu) \sin(\mu)\\
	z &=& -\sin(\mu) \sin(\zeta + \nu) .  \nonumber
\end{eqnarray}
Angles $\mu$ and $\nu$ are clearly polar and azimuthal spherical coordinates, and the lens fraction, $\zeta$, maps out an azimuthal trajectory---i.e. an arc of constant latitude on a circle.

Now apply this reasoning to a cylindrical lens misoriented by an angle of $\eta$ with respect to the y-axis of the pre-lens beam. The incoming beam can be decomposed into active and passive components which are then subjected to the lens operator of Eq. \ref{G}. The resulting spinor can subsequently be converted back to the original basis. With lens orientation described by
\begin{equation}\label{Rot}
	[R_\eta] := \begin{pmatrix} \cos\eta && -\sin\eta \\ \sin\eta && \cos\eta \end{pmatrix} .
\end{equation}
The output spinor is then~\cite{Padgett1999}
\begin{equation}\label{etainout}
	\underline{v}^{\rm (out)} = [R_\eta]^T [G] [R_\eta] \underline{v}^{\rm (in)}.
\end{equation}

\begin{figure}[t]
	\begin{center}
		\includegraphics[width=0.8\linewidth]{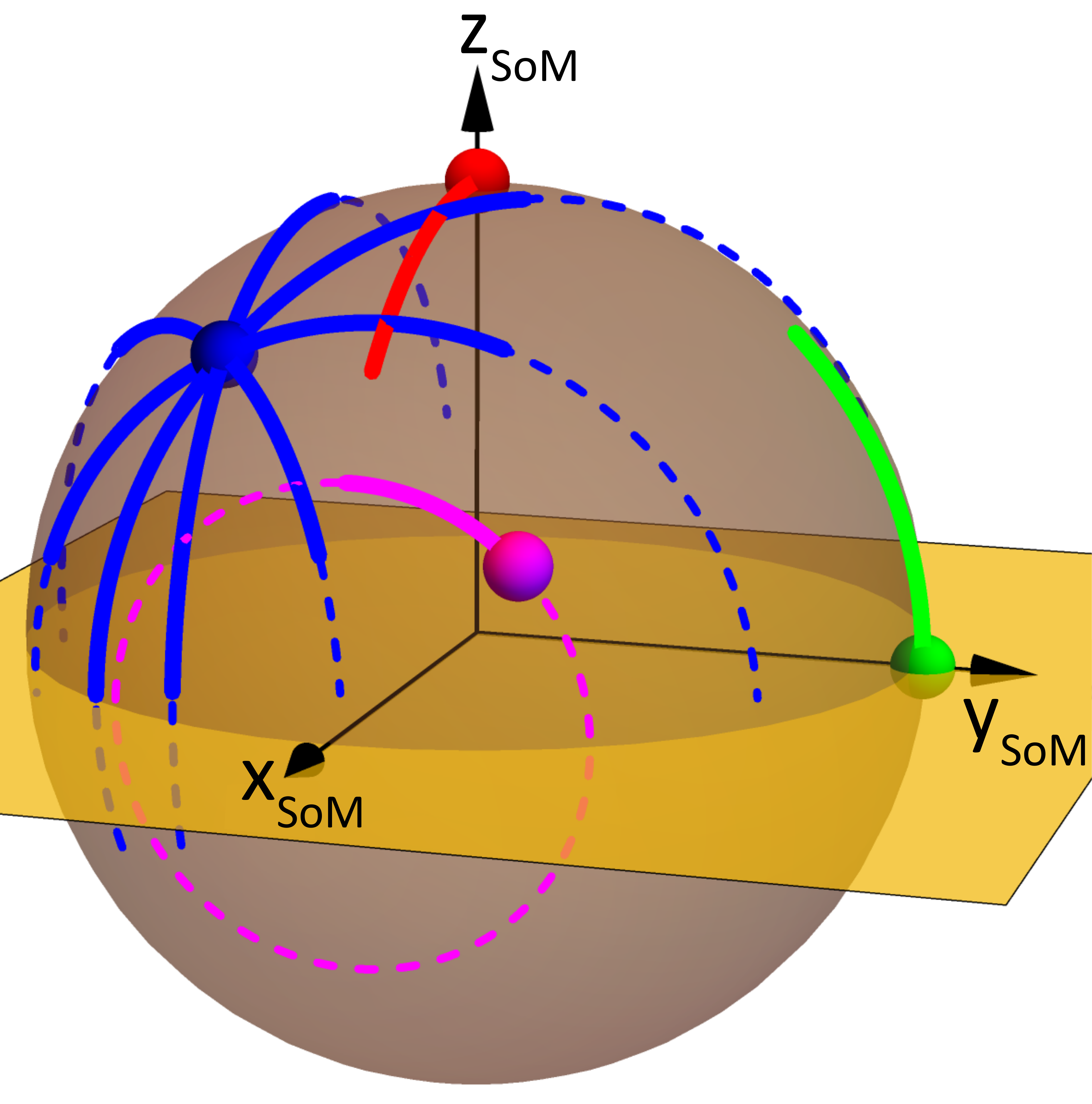}
	\end{center}
	\caption{ \emph{SOM Trajectories for Cylindrical Lens Transits}. Trajectories generated by a lens with $\alpha = \pi/2$ are shown for several sets of initial positions (colors) and lens orientations, $\eta$. The start of each trajectory is indicated with a small sphere of the same color. Dashing is a guide to the eye to more easily see that each trajectory lies on a circle. Red: $\underbar{r}_{\rm init} = \{0,0,1\}$, $\eta =\pi/4$; Green: $\underbar{r}_{\rm init} = \{0,1,0\}$, $\eta =\pi/2$; Magenta: $\underbar{r}_{\rm init} = \{0.861, 0.357,0.362\}$, $\eta =0$; and  Blue: $\underbar{r}_{\rm init} = \{0.612, -0.353,0.707\}$, with $\eta = j \pi/8$, $j\in [0,7]$.}
	\label{PS_Arcs_2}
\end{figure}

Several lens-transit trajectories are plotted in Fig. \ref{PS_Arcs_2} to show how different combinations of input beam and lens orientation map out circular arcs on the SOM. Each arc is obtained by plotting the SOM coordinates as functions of lens transit fraction, $\zeta$.

%
%
%
%

\section{Discussion}
First-order, elliptical, optical vortices accumulate geometric phase as they propagate through astigmatic (cylindrical) lenses. While this has been considered for almost thirty years~\cite{vanenk1993}, a number of outstanding issues have been addressed in this paper by explicitly considering fiber bundles for overall phase and by introducing a Supplementary Product Space that allows beam waists and their positions to be explicitly tracked. Each lens can be decomposed into a stack of differentially thin components, allowing the beam to be described in terms of functions of lens transit fraction. While this decomposition is not unique, it is arguably the simplest and most physically grounded deconstruction of a lens. It also provides a clear explanation for how the mode transits the Sphere of Modes while propagating through the lens.

A prescribed first-order Gaussian beam was applied to two distinctly different cylindrical lenses to show how SOM arcs, fiber position, and location in the SPS can be traced out as functions of transit fraction through each lens. It was shown that identical transits on the SOM can be associated with substantially different input and output beams. 

A projection operator was then constructed that allowed us to determine gauge invariant measures of dynamic phase, total phase, and geometric phase for beam transits through a single lens. The resulting expression for geometric phase was shown to be equal to that obtained using the standard Pancharatnam connection. 

The insights and methodology offered in this work are expected to prove useful in a variety of applications more general than the single-lens focus of the present work. For instance, it is much more common to quantify geometric phase using mode converters that each utilize a pair of lenses. While this makes the measurement of geometric phase more straightforward, it obscures the fact that geometric phase is accumulated as beams propagate through each lens. The approach developed here can be applied equally well to the second lens of a mode converter to determine how its position, relative to the first, influences the associated arc length traced out on the SOM. 

Experimental realization of the predictions of sections I-IV are straightforward since the measurement of lens-modified modes is possible with modern interferometry~\cite{Andersen2019}. Direct measurements of geometric phase provide a greater experimental challenge, though, because of the need to remove dynamic phase contributions~\cite{Bhandari1989,Galvez2003,Maji2019}, and this is even more challenging for non-geodesic trajectories~\cite{Bhandari1989}. A key element of future work will be to experimentally realize the projection operation for tracing out geodesics on the Sphere of Modes while preserving position within the Supplementary Product Space.

The single-lens setting of this work also seems ideal for re-visiting the relationship between geometric phase and the change in orbital angular momentum imposed on the beam by the lens~\cite{vanenk1993}. As a final note, perhaps important in future applications, the framework established in this analysis applies equally well to single-photon settings.

\section{Acknowledgments}
The authors acknowledge useful discussions with P. Ford and support from both the W.M. Keck Foundation and the NSF (DMR 1553905).


\end{document}